\documentclass[10pt]{article} 
\usepackage{wrapfig}
\usepackage{makecell}

\usepackage{soul} 

\usepackage[accepted]{rlc}

\usepackage{amssymb}            
\usepackage{mathtools}          
\usepackage{mathrsfs}           
\mathtoolsset{showonlyrefs}     
\usepackage{graphicx}           
\usepackage{subcaption}         
\usepackage[space]{grffile}     
\usepackage{url}                

\usepackage{xcolor}
\usepackage{soul}
\usepackage{float}
\usepackage{rotating}

\usepackage{algorithm}
\usepackage{algorithmic}
\usepackage{amsmath}

\title{Assigning Credit with Partial Reward Decoupling in Multi-Agent Proximal Policy Optimization}

\author{Aditya Kapoor  \\
    Research \& Innovation, \\
    Tata Consultancy Services,\\
    Mumbai
    \And
    Benjamin Freed \\
    Robotics Institute, \\ 
    Carnegie Mellon University, \\
    Pittsburgh, PA
    \And
    Jeff Schneider \\
    Robotics Institute, \\
    Carnegie Mellon University, \\
    Pittsburgh, PA
    \And
    Howie Choset \\
    Robotics Institute, \\
    Carnegie Mellon University, \\
    Pittsburgh, PA}

\begin{document}

\maketitle

\begin{abstract}
Multi-agent proximal policy optimization (MAPPO) has recently demonstrated state-of-the-art performance on challenging multi-agent reinforcement learning tasks.  However, MAPPO still struggles with the credit assignment problem, wherein the sheer difficulty in ascribing credit to individual agents' actions scales poorly with team size.  In this paper, we propose a multi-agent reinforcement learning algorithm that adapts recent developments in credit assignment to improve upon MAPPO. Our approach leverages partial reward decoupling (PRD), which uses a learned attention mechanism to estimate which of a particular agent's teammates are relevant to its learning updates. We use this estimate to dynamically decompose large groups of agents into smaller, more manageable subgroups. We empirically demonstrate that our approach, PRD-MAPPO, decouples agents from teammates that do not influence their expected future reward, thereby streamlining credit assignment. We additionally show that PRD-MAPPO yields significantly higher data efficiency and asymptotic performance compared to both MAPPO and other state-of-the-art methods across several multi-agent tasks, including StarCraft II.  Finally, we propose a version of PRD-MAPPO that is applicable to \textit{shared} reward settings, where PRD was previously not applicable, and empirically show that this also leads to performance improvements over MAPPO.

\end{abstract}

\section{Introduction}


Multi-agent reinforcement learning (MARL) has achieved super-human performance on many complex sequential decision-making problems, such as DOTA 2 \citep{openai5}, StarCraft II \citep{starcraft}, and capture the flag \citep{capture_the_flag}.
These impressive results, however, come at an immense cost: often, they require millions, if not billions, of time-consuming environmental interactions, and therefore can only be run on high-cost compute clusters.

The \textit{credit assignment problem} contributes to the computational difficulties that plague large-scale MARL algorithms; as the number of agents involved in learning increases, so too does the difficulty of assessing any individual agent's contribution to overall group success \citep{minsky,sutton_barto}. While credit assignment already challenges reinforcement learning (RL), it is particularly prominent in large-scale \textit{cooperative} MARL, because, unlike problems in which each agent can act greedily to optimize its own reward, all agents must maximize the total reward earned by the entire group. Therefore, agents must not only consider how their actions influence their own rewards, but also the rewards of every other agent in the group.

A popular class of approaches to MARL are policy-gradient methods, which also suffer from the credit assignment problem.  Recent work in improving policy-gradient methods took the approach of developing concepts which were then used to extend the original actor-critic algorithm.  These extensions include counterfactual multi-agent policy gradients (COMA) \citep{coma}, multi-agent game abstraction via graph attention neural networks (G2ANet) \citep{g2anet}, and partial reward decoupling (PRD) \citep{prd}. \textbf{The primary contributions of this paper are 1) the machinery necessary for applying PRD to a state-of-the-art multi-agent policy-gradient method (multi-agent PPO (MAPPO)), and 2) a version of PRD that does not require the environment to provide individual rewards streams for each agent, and instead utilizes a \textit{shared} reward signal.}

PRD simplifies credit assignment by decomposing large cooperative multi-agent problems into smaller decoupled subproblems involving subsets of agents. PRD was applied to the actor-critic algorithm \citep{prd, actor-critic}. Meanwhile, significant progress has been made towards improving the data efficiency of policy-gradient algorithms. Most notably, trust-region policy optimization (TRPO) and proximal policy optimization (PPO) improve the data efficiency of actor-critic algorithms by enabling a given batch of on-policy data to be re-used for multiple gradient updates. PPO, in particular, has demonstrated strong performance in multi-agent settings \citep{ma_ppo}. However, we argue that because PPO relies on stochastic advantage estimates, it still suffers from the credit assignment problem, and can therefore be improved by incorporating advanced credit assignment strategies.  

In this paper, we demonstrate that PRD can be leveraged within the learning updates of PPO for each individual agent, to eliminate the contributions from other irrelevant agents.  We find that the resulting algorithm, PRD multi-agent PPO (PRD-MAPPO), exceeds the performance of prior state-of-the-art MARL algorithms such as QMix \citep{qmix}, MAPPO \citep{ma_ppo}, LICA \citep{lica}, G2ANet \citep{g2anet}, HAPPO \citep{happo} and COMA \citep{coma} on a range of multi-agent benchmarks, including StarCraft II.
Beyond integrating PRD with MAPPO, we make three key modifications to the original PRD approach proposed by \cite{prd}.  First, we introduce a ``soft'' variant that softly re-weights advantage terms in agents' learning updates based on attention weights, rather than the strict decoupling used by \cite{prd}.  Second, we modify the advantage estimation strategy that allows learning updates to be computed in time that is linear, rather than quadratic, in the number of agents.  Finally, we propose a version of PRD-MAPPO that is capable of using \textit{shared} rewards, as opposed to individual agent rewards, thus broadening the range of problems to which our algorithm can be applied.

To gain deeper insight to the source of PRD-MAPPO's improved performance, we visualize the relevant sets identified by PRD, and verify that PRD decomposes multi-agent teams into subsets of agents that should cooperate with one another. Finally, we compare the gradient estimator variance of PRD-MAPPO and MAPPO, and find that PRD-MAPPO indeed tends to avoid the spikes in gradient variance present in MAPPO, helping explain its superior data efficiency and stability.

\section{Background}
Here we describe our problem formulation as a Markov game.
Subsequently, we investigate mathematically how imperfect credit assignment manifests itself in high policy-gradient variance in policy-gradient RL algorithms.
Finally, we review PPO and PRD.

\subsection{Markov Games}
\label{subsec:mmdp}
We consider multi-agent sequential decision-making problems that can be modeled as a Markov game. A Markov game is specified by \( (\mathcal{S},\mathcal{A},\mathcal{P},\mathcal{R},\rho_0,\gamma ) \), where \(\mathcal{S}\) is the state space, \(\mathcal{A}\) is the joint action space, consisting of every possible combination of individual agents' actions, \(\mathcal{P}(s_{t+1}|s_t,a_t)\) specifies the state transition probability distribution, \(\mathcal{R}(r_t|s_t,a_t)\) specifies the reward distribution, \(\rho_0(s_0)\) denotes the initial state distribution, and \( \gamma \in (0,1]\) denotes a discount factor \citep{markov_game}. 
At each timestep \(t \in \{0,...,T\}\), each agent \(i \in \{1,...,M\}\) selects an action independently according to its state-conditioned policy \( \pi_i(a_t^{(i)}|s^{(i)}_t;\theta_i) \).  Here, \(T\) specifies the episode length, \(M\) denotes the number of agents, \(s^{(i)}_t\) denotes the state information available to agent \(i\), and \( \theta_i \) denotes the parameters for agent \(i\). Subsequently, individual agent rewards are sampled according to \(r^{(1)}_t,...,r^{(M)}_t \sim \mathcal{R}(\cdot|s_t,a_t)\), and the state transitions according to \(s_{t+1} \sim \mathcal{P}(\cdot|s_t,a_t)\).

Although agents receive individual rewards, we are primarily interested in learning \textit{cooperative} behaviors that maximize \textit{total} group return, that is, the sum of all agents' individual rewards across all timesteps. More precisely, we wish to find the optimal agent policy parameters \(\theta^* = \{\theta_1^*,...,\theta_M^*\} = \underset{\theta}{\mathrm{argmax}}  J(\theta) \), where
\begin{equation}
    \label{eq:obj}
    J(\theta) = \mathbb{E} \Big[ \sum_{t=0}^T \sum_{j=1}^M \gamma^t r_t^{(j)} \Big| \pi_{\theta} \Big].
\end{equation}  

This problem formulation is distinct from the ``greedy'' case, where each agent maximizes its own individual return.  In this problem formulation, agents should learn to be altruistic in certain situations, by selecting actions that help maximizes group reward, possibly at the expense of some individual reward.





\subsection{Credit Assignment and Policy Gradient Variance}
\label{subsec:credit_assignment_variance}

To understand the effects of scaling PPO to large numbers of agents, and how we expect PRD will improve this scaling, we explore how imperfect credit assignment causes difficulties in learning. In this paper, we argue that in policy-gradient algorithms (which includes many popular algorithms such as PPO \citep{ppo}, TRPO \citep{trpo}, D4PG \citep{d4pg}, MADDPG \citep{multi-agent_ac}, and A3C \citep{a3c}), the credit assignment problem manifests itself in the form of high variance of advantage estimates.  High variance in advantage estimates in turn causes policy gradient estimates to be more noisy, resulting in slower learning.  

We consider an actor-critic-style gradient estimate for a single-agent system in its most stripped-down possible form, computed using a single state-action sample:

\begin{equation}
    \hat{\nabla}_{\theta} J(\theta,s,a) = \nabla_{\theta} \log \pi(a|s) \hat{A}(s,a),
\end{equation}

\noindent where state \(s\) is sampled from the state-visitation distribution induced by policy \(\pi\), action \(a\) is sampled from \(\pi\) conditioned on \(s\), and \(\hat{A}(s,a)\) is a \textit{stochastic advantage estimate}, which estimates the true \textit{advantage} of taking action \(a_t\) in state \(s_t\), and following policy \(\pi\).  The advantage function is typically defined as \(A^{\pi}(s,a) = Q^{\pi}(s,a) - V^{\pi}(s)\), where \(Q^{\pi}(s,a)\) and \(V^{\pi}(s)\) are the state-action value function and state-value function, respectively \citep{sutton_barto}. Intuitively, the advantage function measures how much better it is to select a particular action \(a\) than a random action from the policy, while in state \(s\). There are many ways to compute \(\hat{A}\), generally all involving some error, as the true value functions are unknown  \citep{sutton_barto,gae}. If perfect advantage estimation were possible, then so too would be perfect credit assignment, as the advantage function directly measures how a particular action \(a\) impacted the total reward obtained by the group.  

To gain an understanding of how the gradient variance is impacted by advantage estimator variance, we note that the conditional variance of \(\hat{\nabla}_{\theta} J\), given \(s\) and \(a\), is proportional to the variance of \(\hat{A}\):

\begin{equation}
    \mathrm{Var}(\hat{\nabla}_{\theta} J|s,a) = \left( \nabla_{\theta} \log \pi(a|s) \right)\left( \nabla_{\theta} \log \pi(a|s) \right)^T \mathrm{Var}(\hat{A}|s,a).
\end{equation}

Moving to a cooperative multi-agent setting, \(\hat{A}(s,a)\) is replaced by a summation over individual agents' advantages in the gradient estimate for a particular agent \(i\):

\begin{equation}
    \hat{\nabla}_{\theta_i} J(\theta,s,a) = \nabla_{\theta_i} \log \pi_i(a_i|s) \sum_{j=1}^M \hat{A}_{ij}(s,a),
\end{equation}

\noindent where \(\hat{A}_{ij}(s,a)\) now corresponds to our estimate of how agent \(i\)'s action influenced the expected future reward of agent \(j\). The summation results from the fact that in the cooperative setting, agent \(i\) is no longer interested only in maximizing its own total reward, but is instead interested in maximizing \textit{total group reward}, as discussed in Sec.~\ref{subsec:mmdp}. The variance of \(\hat{\nabla}_{\theta_i} J\) given \(s\) and \(a\) now depends on the variance of each individual agent's advantage estimates, as well as the covariance between every pair of agents' advantages. Using Bienaymé's identity, and omitting the arguments to \(\pi_i\) for brevity, we can express this variance as 
    \begin{equation}
         \mathrm{Var}(\hat{\nabla}_{\theta_i} J|s,a) =  \left( \nabla_{\theta_i} \log \pi_i \right)  \left( \nabla_{\theta_i} \log \pi_i \right)^T \Bigg( \sum_{j=1}^M \mathrm{Var}(\hat{A}_{ij}|s,a) \\+ 2 \sum_{k<j} \mathrm{Cov}(\hat{A}_{ij},\hat{A}_{ik}|s,a) \Bigg).
    \end{equation}

To simplify analysis, we consider an upper bound on gradient estimator variance, obtained using the Cauchy–Schwarz inequality,

    

\begin{equation}
\label{eq:variance}
    \mathrm{Var}(\hat{\nabla}_{\theta_i} J|s,a) \leq  \left( \nabla_{\theta_i} \log \pi_i \right)  \left( \nabla_{\theta_i} \log \pi_i \right)^T \Bigg( \sum_{j=1}^M \mathrm{Var}(\hat{A}_{ij}|s,a) \\+ 2 \sum_{k<j} \sqrt{\mathrm{Var}(\hat{A}_{ij}|s,a) \mathrm{Var}(\hat{A}_{ik}|s,a)} \Bigg),
\end{equation}

\noindent which can be seen to scale roughly linearly with number of agents, assuming \(\mathrm{Var}(\hat{A}_{ij}|s,a)\) is roughly similar for all \(j\). Therefore, to achieve a particular signal-to-noise ratio, more such gradient estimates will need to be averaged together as team size increases, thus increasing the data requirements of the algorithm. This analysis helps explain the mechanism by which improved credit assignment can yield data-efficiency improvements for policy-gradient algorithms, such as A3C \citep{a3c}, TRPO \citep{trpo} and PPO \citep{ppo} algorithms. In particular, our approach aims to eliminate extraneous advantage terms that do not on average contribute to the policy gradient, thereby reducing the number of terms in the summations in \eqref{eq:variance} and decreasing the total variance.  We discuss this further in Sec. \ref{sec:prd} and \ref{sec:prd-ppo}.

\subsection{Proximal Policy Optimization} 
\label{sec:ppo}

Earlier policy gradient algorithms, such as actor-critic (AC), suffered from poor data efficiency in part because they were purely on-policy, and therefore required a fresh batch of environmental data to be collected each time a single gradient update was applied to the policy \citep{actor-critic,trpo,ppo}. PPO provides higher data efficiency than AC by enabling multiple policy updates to be performed given a single batch of on-policy data, resulting in larger policy improvements for a fixed amount of data. Given a batch of data, PPO optimizes the policy with respect to a ``surrogate'' objective that penalizes excessively large changes from the old policy, permitting the agent to perform multiple gradient updates without becoming overly off-policy. Specifically, during each policy optimization step, PPO optimizes the following objective with respect to policy parameters \(\theta\),

\begin{equation}
    \label{eq:ppo_obj}
    L_{\mathrm{PPO}}(\theta) = \hat{\mathbb{E}} \left[ \mathrm{min} \left( \big( r_t(\theta) \hat{A}_t \big), \big( \mathrm{clip}(r_t(\theta),1-\epsilon,1+\epsilon) \hat{A}_t \big) \right) \right],
\end{equation}

\noindent where \(r(\theta)=\frac{\pi_{\theta}(a_t|s_t)}{\pi_{\theta_{old}}(a_t|s_t)}\) is the probability ratio, \(\pi_{old}\) is the data collection policy, \(\pi\) is the updated policy, \(\hat{A}_t\) is the stochastic advantage estimate for time \(t\), and \(\hat{\mathbb{E}}[\cdot]\) denotes an empirical average over a finite batch of samples \citep{ppo}.

PPO has been recently shown to offer strong performance on multi-agent problems \citep{ma_ppo}. However, PPO does not explicitly control the variance of its policy gradient updates, which as we discuss in Sec. \ref{subsec:credit_assignment_variance}, tends to grow with multi-agent team size. This increased gradient estimate variance means that larger batches of data become necessary to reach a satisfactory signal-to-noise ratio in the learning updates; indeed, \citep{ma_ppo} found that much larger batch sizes were necessary for PPO to perform well on multi-agent tasks.  In this work, we seek to combine the data efficiency benefits of PPO with the variance reduction benefits of PRD, to enable further improvements in data efficiency and stability.

\subsection{Partial Reward Decoupling}
\label{sec:prd}
PRD is an approach that enables large multi-agent problems to be dynamically decomposed into smaller subgroups such that cooperation among subgroups yields a fully cooperative group-level solution. In practice, PRD was shown to improve the performance of an AC-style approach, compared to a vanilla AC algorithm. The proposed PRD-AC algorithm was also shown to outperform COMA, a popular method for improved multi-agent credit assignment.  




PRD makes use of the fact that, considering two agents \(i\) and \(j\) at a particular timestep \(t\), if the action of agent \(i\) does not influence the expected future reward of agent \(j\), then agent \(i\) need not take agent \(j\)'s rewards into account when computing its advantage estimate for time \(t\), thus streamlining credit assignment. The set of agents whose expected future rewards are impacted by the action of agent \(i\) at time \(t\) is referred to as the \textit{relevant set} of agent \(i\) at time \(t\), denoted \(R^{\pi}_i(s_t,a_t)\).  In \cite{prd}, a learned value function with an attention mechanism was used to estimate the relevant set for each agent.  

There were significant drawbacks to the approach presented by \cite{prd}, which we address in this paper. First, PRD was used in the context of the AC algorithm, which has been surpassed by algorithms such as TRPO and PPO. Second, for a problem involving \(M\) agents, PRD required \(M\) evaluations of the critic function to compute a single agent's gradient update; thus the computational burden for a learning update scaled quadratically with the number of agents. 
Finally, PRD assumed that the environment provided per-agent reward streams (\textit{i.e.}, provided a scalar reward value to each agent at each timestep).  However, many multi-agent problems provide only a single scalar reward for the entire group at each timestep.


\section{Improving Proximal Policy Optimization with Partial Reward Decoupling}
\label{sec:prd-ppo}

In this paper, we tackle the credit assignment problem by developing PRD-MAPPO, which leverages a PRD-style decomposition within a PPO learning update to improve credit assignment. More specifically, PRD modifies the original PPO objective by eliminating advantage terms belonging to ``irrelevant'' agents. As shown by \cite{prd}, these irrelevant advantage terms contribute only noise to learning updates, making learning less efficient.  PRD uses an attention-based value function to identify when a particular agent's action did not influence another agent's future return, allowing those agents to be decoupled.


To leverage the improved credit assignment capabilities of PRD in PPO, we make two modifications to the standard PPO algorithm: first, we incorporate a learned critic with an attention mechanism. Similar to \cite{prd}, the attention weights computed by the critic will be used to estimate the relevant set of agents, as described in Sec. \ref{subsec:critic}. Unlike \cite{prd}, we modify the critic architecture to allow the relevant sets for each agent to be computed in linear, rather than quadratic time. Second, we modify the surrogate objective of PPO to use the streamlined advantage estimation strategy of PRD, which we describe in Sec. \ref{subsec:param_update}, using the relevant set estimated using the critic. In this work, we test a novel ``soft'' relevant set estimation strategy that softly decouples agents, which we find significantly improves performance over a manual thresholding approach as was used by \cite{prd}. 

\subsection{Learned Critics for Relevant Set and Advantage Estimation}
\label{subsec:critic}

\begin{wrapfigure}{R}{0.55\textwidth}
    \vspace{-5mm}
    \centering
    \includegraphics[width=.55\textwidth]{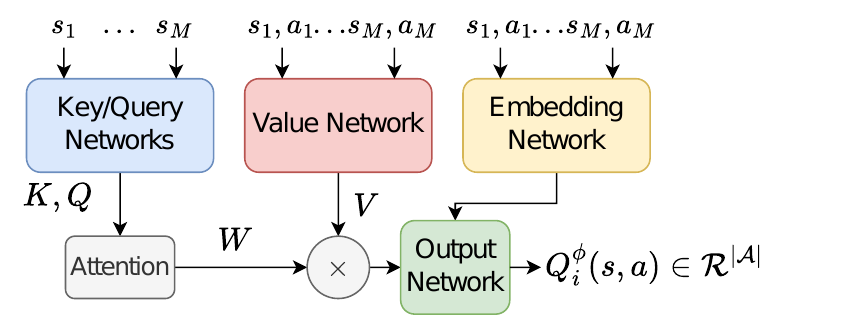}
     \caption{Q and Value Function Network Architecture.  Each agent uses states from all agents to compute attention weights for every agent other than itself. These attention weights are then used to aggregate attention values from all agents other than itself. Finally, aggregated attention values for agent \(i\) are concatenated either with the embedded state-action vector for agent \(i\) (if the network is functioning as a Q function) or the embedded state vector for agent \(i\), (if the network is functioning as a value function). Finally, this is passed through the output network to generate either \(Q^{\phi}_i(s,a)\) or \(V^{\psi}_i(s,a^{\neq i})\).} 
    \label{fig:critic}
    \vspace{-4mm}
\end{wrapfigure}

Similar to \cite{prd}, we use a learned critic function to perform relevant set estimation, albeit with significant modifications. In our approach, each agent \(i\) maintains a graph neural network Q function \(Q_i^{\phi}(s_t,a_t)\), which is trained to estimate its expected future individual returns given the current state and actions of all agents.  A diagram of our Q function is depicted in Fig. \ref{fig:critic}. In practice, all agents share the same Q function parameters. 
\(Q_i^{\phi}\) takes as input the state information and actions of all agents to estimate a scalar Q value for each agent \(i\).

The Q function contains an attention mechanism that allows it to ``shut off'' dependence on particular agents' actions. More concretely, the Q network for each agent \(i\) uses the states of all agents (including itself) to compute attention weights for all other agents (agents assign an attention weight of 1 to themselves, \textit{i.e.}, \(w_{ii}(s_t)=1\)). These attention weights are then used as coefficients to compute a linear combination of attention values computed from agents' states and actions.  If a particular attention weight \(w_{ij}\) is \(0\), then any information about agent \(j\)'s action will not be propagated further through the network, meaning that agent \(j\)'s action will not influence the final Q estimate for agent \(i\).  Once the aggregated value is computed, it is concatenated with an embedding computed from agent \(i\)'s state and action and passed through a recurrent output network (Fig.~\ref{fig:critic}).   

If the learned Q function of agent \(i\) at a particular timestep \(t\) computes an attention weight of exactly zero for another agent \(j\) (\textit{i.e.,} \(w_{ij}(s_t)=0\)), then \(Q_i^{\phi}\) does not depend on \(a^{(j)}_t\) given the state of all agents, and we can infer that agent \(i\) is outside the relevant set of agent \(j\). As shown by \cite{prd}, agents outside the relevant set of agent \(j\) do not, on average, contribute to its policy gradient, and may therefore be removed from the policy gradient estimates without introducing bias.   In practice, when inferring the relevant sets for each agent, we infer that \(i \notin R_j(s_t)\) if \(w_{ij}(s_t) < \epsilon\), where \(\epsilon > 0\) is a small manually chosen constant.  Using this soft attention mechanism, agents cannot assign precisely zero attention weight to any other agent, and therefore cannot guarantee complete independence of the Q function to any particular agent's action.  However, we found that in practice, very small attention weights were assigned to irrelevant agents, making this a practical method for relevant set estimation.  We explore variants of this decoupling procedure, including a ``soft'' variant that softly re-weights agents' contributions to learning updates. 


Our approach to computing advantage terms for learning updates reduces the computational complexity over \citep{prd} from quadratic to linear in the number of agents \(M\).
To compute the advantage terms required to update the policy of a particular agent \(i\), the original algorithm described by \cite{prd} requires each agent \(i\) to estimate the expected future return of each agent \(j\), conditioned on the actions of all agents other than \(i\), for each \(j \in R_i(s_t)\). 
This computation requires (at worst) \(M\) calls to the critic for each of the \(M\) agents, resulting  \(M^2\) total calls during each learning update. 
Our approach, on the other hand, circumvents with quadratic scaling by maintaining two separate critics; the first is the Q function used for relevant set estimation, described above.  The second critic is used solely to provide baseline estimates for advantage function estimation \citep{gae,actor-critic}.
It estimates the \textit{sum} of expected future returns for all agents within agent \(i\)'s relevant set, conditioned on the state of all agents, and the actions of all agents other than \(i\).  We refer to this critic as the \textit{value} function, rather than the Q function, because it does not depend on the actions of agent \(i\).  The value function uses an architecture almost identical to the Q function (Fig. \ref{fig:critic}), with the one difference that the attention values are concatenated with the embedded \textit{state} of agent \(i\), rather than state-action.  Using this value function, computing advantages for all agents requires only \(M\) calls (one  per agent).

\subsection{PRD-MAPPO Parameter Update Rule}
\label{subsec:param_update}

We modify the original MAPPO \citep{ma_ppo} objective for each agent \(i\) by eliminating the rewards from agents that are outside its relevant set from its advantage estimates.  
The original MAPPO algorithm optimizes the following objective during each policy parameter update for agent \(i\):

\begin{equation}
\label{eq:mappo_obj}
    L^{(i)}_{\mathrm{MAPPO}} = \hat{\mathbb{E}} \Bigg[ \mathrm{min}  \Bigg(  \Big( r^{(i)}_t(\theta_i) \hat{A}_{t} \Big),  \Big(\mathrm{clip}(r_t^{(i)}(\theta_i),1-\epsilon,1+\epsilon) \hat{A}_{t}\Big) \Bigg) \Bigg],
\end{equation}

\noindent where \(r^{(i)}\) is the ratio between the updated and old policy of agent \(i\), and \(\hat{A}_{t}\) is the advantage estimate for timestep \(t\).  In  \citep{ma_ppo}, generalized advantage estiamtion was used to compute \(\hat{A}_{t}\), which combines group agent rewards and value function estimates according to

\begin{align}
    & \hat{A}_t = \delta_t + (\gamma \lambda) \delta_{t+1} + ... + (\gamma \lambda)^{T-t+1} \delta_{T-1}, \\
   & \mathrm{where} \quad \delta_t= \left( \sum_{j=1}^M r^{(j)}_t \right) + \gamma V^{\psi}(s_{t+1}) - V^{\psi}(s_{t}).
\end{align}

We modify the objective in \eqref{eq:mappo_obj} by replacing advantage terms with \textit{individual agent} advantage terms, which ignore the rewards of irrelevant agents.  The objective for agent \(i\) becomes 


\begin{equation}
\label{eq:prd_mappo}
    L^{(i)}_{\mathrm{PRD}} = \hat{\mathbb{E}} \Bigg[ \mathrm{min}  \Bigg(  \Big( r^{(i)}_t(\theta_i) \hat{A}_{i,t} \Big),  \Big(\mathrm{clip}(r_t^{(i)}(\theta_i),1-\epsilon,1+\epsilon)  \hat{A}_{i,t} \Big) \Bigg) \Bigg],
\end{equation}

\noindent where 
\begin{align}
    \hat{A}_{i,t} &=  \delta_{i,t} + (\gamma \lambda) \delta_{i,t+1} + ... + (\gamma \lambda)^{T-t+1} \delta_{i,T-1}, \\
     \delta_{i,t} &=  \left( \sum_{j \in R_i(s_t)} r_t^{(j)} \right) + \gamma V_i^{\psi}(s_{t+1},a_{t+1}^{\neq i}) - V_i^{\psi}(s_{t},a_{t}^{\neq i}). 
\end{align}

Note in the above equation that the reward terms for agents not in \(R_i(s_t)\) have been removed, and \(V^{\psi}\) has been replaced by the value function \(V_i^{\psi}\) described in Sec. \ref{subsec:critic}., which is regressed against the sum of returns of agents in \(R_i(s_t)\).  Pseudocode for PRD-MAPPO is included in Sec.~\ref{sec:pseudocode} of the appendix.

We additionally propose a ``soft'' variant of PRD-MAPPO, which we refer to as PRD-MAPPO-soft, that softly reweights agent rewards according to attention weights of the Q network, \textit{i.e.}, \(\delta_{i,t} =  \left( \sum_{j=1}^M w_{ji}(s_t) r_t^{(j)} \right) + \gamma V_i^{\psi}(s_{t+1},a_{t+1}^{\neq i}) - V_i^{\psi}(s_{t},a_{t}^{\neq i})\).  In this soft variant, \(V_i^{\psi}\) is regressed against the \textit{weighted sum} of agent returns,  \(\sum_{j=1}^M w_{ji}(s_t) R_t^{(j)} \).

\subsection{Partial Reward Decoupling for environments with shared rewards}
\label{subsec:prd_shared}
One drawback to our PRD approach is that it assumes individual reward streams for each agent are available, \textit{i.e.,} at each timestep, the environment provides a separate scalar reward for each agent.
However, some multi-agent systems only provide a single scalar \textit{shared} reward for the entire group at each timestep.
To deal with the shared reward setting, we propose a strategy for decomposing shared returns into individual agent returns, to which we can then apply PRD.
We start by training a \textit{shared} Q function to predict the shared returns (\textit{i.e.,} the sum of future shared rewards).  
Here we use a similar architecture as described in Sec. \ref{subsec:critic}, with the one difference that our network has 1 output rather than \(M\) outputs.
We denote the vector of attention weights assigned by all agents to the action of agent \(j\) as \(W_{:j}\).
There is one such vector for each timestep and each agent; we omit the timestep subscripting for brevity.
As a heuristic to measure the overall influence that each agent \(j\) has on the future shared reward, we aggregate the attention weights for each agent \(j\) by taking the mean of \(W_{:j}\), which we refer to as \(\Tilde{W}_j\). 
The individual returns for each agent \(j\) at each timestep are then set proportionally to \(\Tilde{W}_j\), such that they sum to the original shared return. 
Subsequently, we apply PRD-MAPPO to these individual returns as we would in the individual reward setting described in Sec. \ref{subsec:param_update}. 
We refer to this approach as PRD-MAPPO-shared.

\section{Experiments}

We experimentally compare the performance of the following algorithms on several cooperative MARL environments: 

\textbf{PRD-MAPPO (ours)}: MAPPO with PRD, as described in Sec. \ref{subsec:critic}.

\textbf{PRD-MAPPO-soft (ours) }: the soft variant of PRD-MAPPO as described in Sec. \ref{subsec:critic}.

\textbf{PRD-MAPPO-shared (ours) }: the soft variant of PRD-MAPPO in the shared reward setting, as described in Sec. \ref{subsec:prd_shared}.


\textbf{MAPPO}: a multi-agent variant of PPO, proposed by \cite{ma_ppo}.


\textbf{HAPPO}: a recent state-of-the-art algorithm proposed by \cite{happo} that extends trust region learning to cooperative multi-agent reinforcement learning (MARL), enabling monotonic policy improvement without the need for shared policy parameters.

\textbf{G2ANet-MAPPO}: MAPPO with a G2ANet-style critic.  This baseline attempts to import the credit assignment benefits of G2ANet (which was originally used in the Actor-Critic algorithm) to the more state-of-the-art MAPPO. 

\textbf{Counterfactual Multi-Agent Policy Gradient (COMA)}: Proposed by~\cite{coma}, COMA is a multi-agent actor-critic method. COMA addresses credit assignment by using a counterfactual baseline that marginalizes out a single agent’s action, while keeping the other agents’ actions fixed, allowing COMA to better isolate each agent's contribution to group reward.

\textbf{PRD-V-MAPPO}: PRD-MAPPO, using the value-function-based method of relevant set estimation, as described by \cite{prd}.  This version uses a learned value function for both relevant set and advantage estimation, and scales quadratically in time complexity with number of agents.  We include this as a baseline to assess the effect of critic choice.

\textbf{Learning Implicit Credit Assignment (LICA)}: proposed by~\cite{zhou2020learning}, LICA is a method for implicit credit assignment that is closely related to value gradient methods, which seek to optimize policies in the direction of approximate value gradients.  LICA extends the concept of value mixing present for credit assignment found in QMix and Value-decomposition Networks by introducing an additional latent state representation into the policy gradients.  The authors claim that this additional state information provides sufficient information for learning optimal cooperative behaviors without explicit credit assignment.

\textbf{QMix:} proposed by \cite{qmix}, QMix learns a joint state-action value function, represented as a complex non-linear combination of per-agent value functions. The joint value function is structurally guaranteed to be monotonic in per-agent values, allowing agents to maximize the joint value function by greedily selecting the best actions according to their own per-agent value functions.


The policy network and critic used for advantage calculations for PRD-MAPPO, PRD-MAPPO-soft, PRD-MAPPO-shared, MAPPO, HAPPO, G2ANet-MAPPO, COMA and PRD-V-MAPPO have the same architecture and number of parameters. Because LICA and QMix depend on a particular critic architecture, we used the original architectures as described by \cite{lica} and \cite{qmix} respectively.
For all environments and all algorithms, we performed a grid search over hyperparameters as described in the appendix.

We consider the following environments, with detailed descriptions of each in the appendix: Collision Avoidance, Pursuit, Pressure Plate, Level-Based Foraging, and StarCraft Multi-Agent Challenge Lite (SMAClite), specifically the 5m\_vs\_6m, 10m\_vs\_11m, and 3s5z battle scenarios.

 \begin{figure}[!ht]
        \centering
        \includegraphics[width=\linewidth]{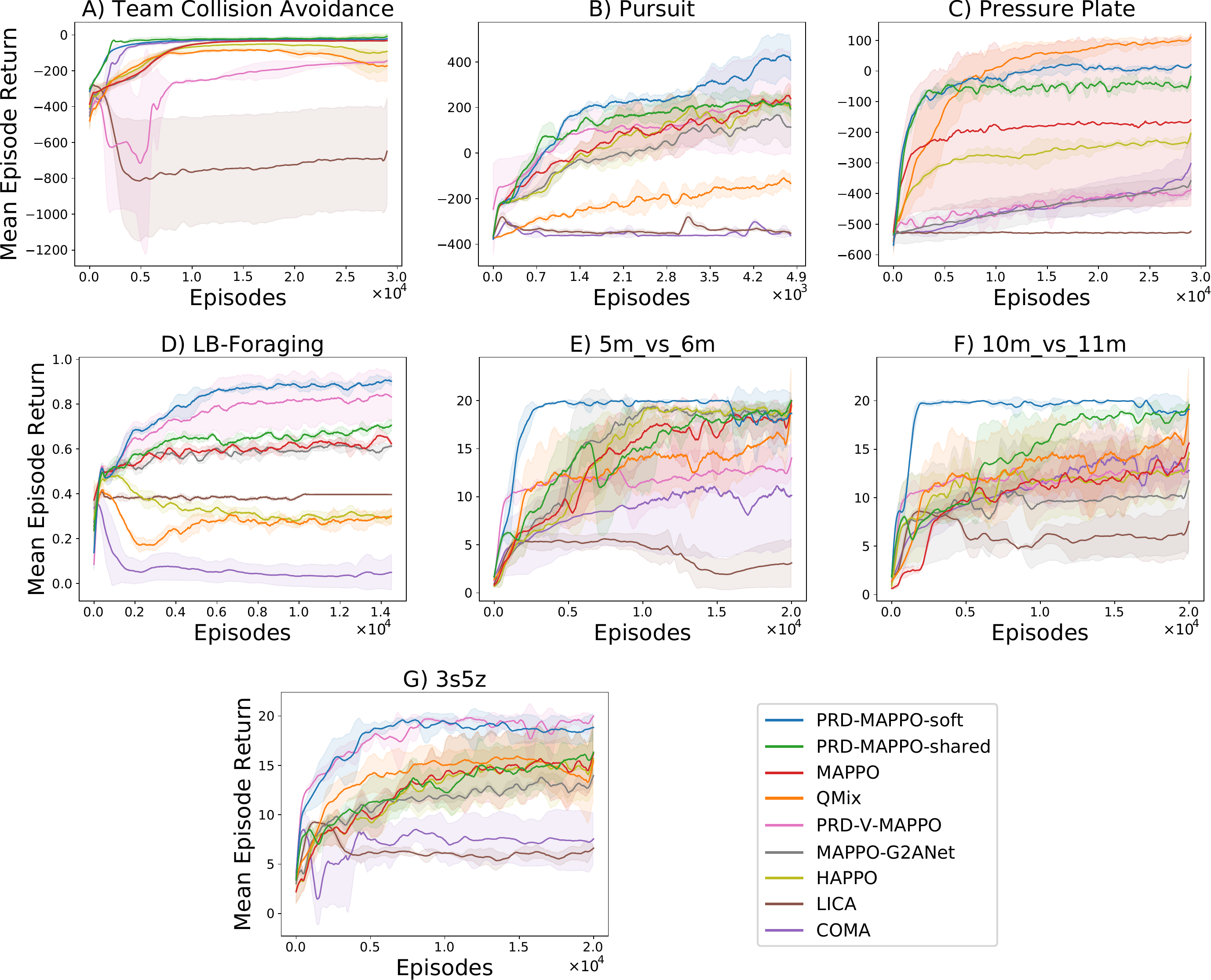}
        \vspace{-0.3cm}
        \caption{\textbf{Average reward vs. episode for PRD-MAPPO-soft, PRD-MAPPO, PRD-V-MAPPO, COMA, LICA, QMix, MAPPO, MAPPO-G2ANet on A) team collision avoidance, B) pursuit, C) pressure plate, D) Level-Based Foraging, E) StarCraft 5m\_vs\_6m, F) StarCraft 10m\_vs\_11m  tasks, and G) StarCraft 3s5v}.  Solid lines indicate the average over 5 random seeds, and shaded regions denote a 95\% confidence interval.  Approaches that incorporate PRD (PRD-MAPPO and PRD-MAPPO-soft) tend to outperform all other approaches, indicating that PRD can be leveraged to improve PPO by improving credit assignment.}
        \label{fig:reward_curves}
        \vspace{-5mm}
\end{figure}

\section{Results and Discussion}

The reward curves for all tasks are shown in Fig. \ref{fig:reward_curves}.  We found that of the algorithms we tested, only PRD-MAPPO-soft, PRD-MAPPO-shared, and PRD-MAPPO performed consistently well across all environments, with PRD-MAPPO-soft tending to perform the best.  PRD-MAPPO-soft was outperformed only in one environment (pressure plate) by one algorithm (QMix), and in general outperformed all other algorithms on all tasks.

\subsection{Relevant Set Visualization}

\begin{wrapfigure}{R}{0.45\textwidth}
    \vspace{-5mm}
    \centering
    \includegraphics[width=.45\textwidth]{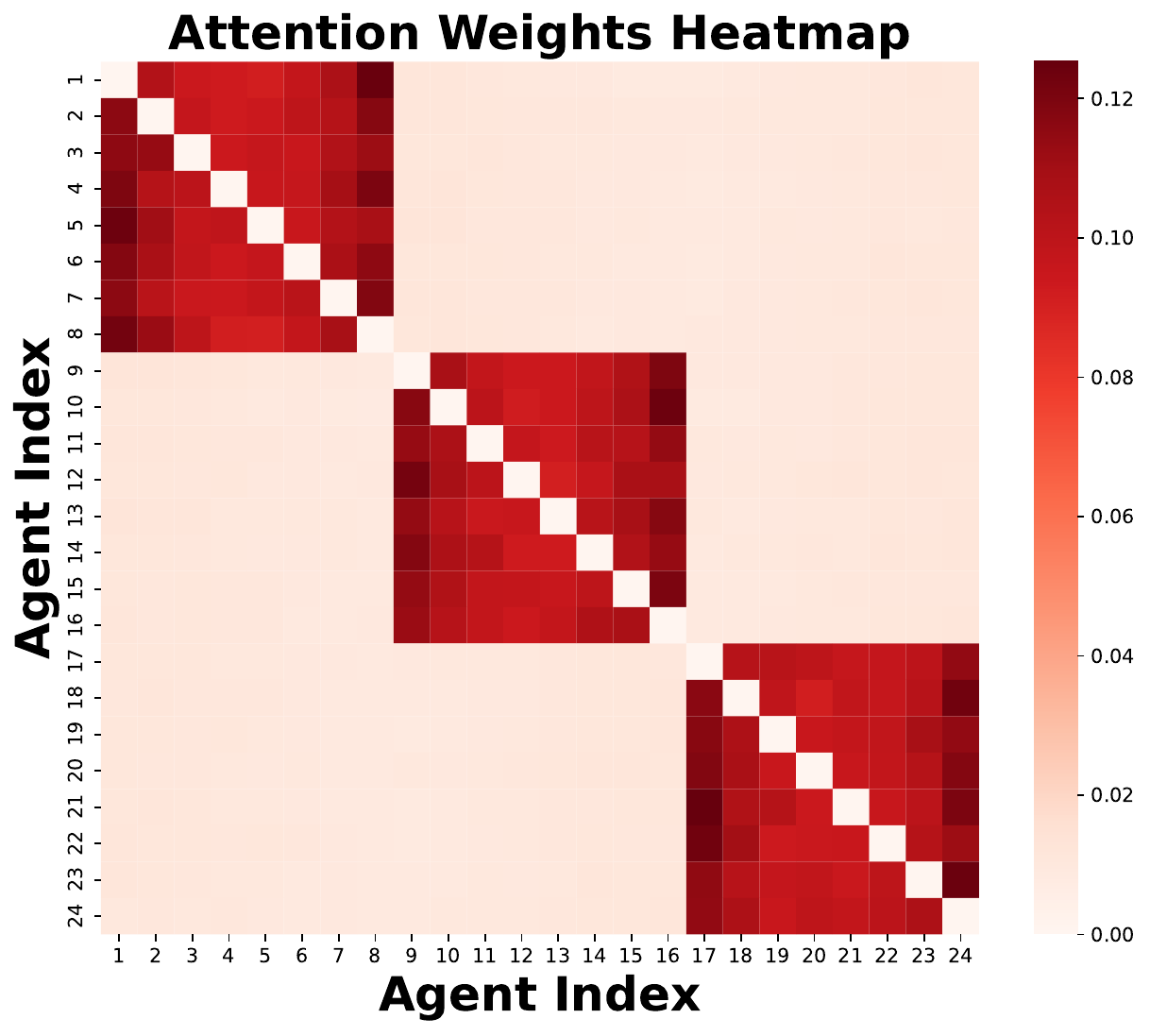}
        \vspace{-6mm}
        \caption{\textbf{Relevant set visualization in Collision Avoidance environment.} We visualize the average attention weight that each agent assigns to every other agent, averaged across 5000 independent episodes.  Because agents always assign an attention weight of 1 to themselves, we remove those elements from the plot as they are uninformative.  We notice that generally agents assign a far higher attention weight to agents in their team, compared to agents on other teams, which is to be expected given that only an agent's teammates are capable of influencing its rewards.}
        \label{fig:relevant_set_viz}
    \vspace{-15mm}
\end{wrapfigure}

To gain more insight into the relevant set selection process, in Fig.~\ref{fig:relevant_set_viz} we visualized the attention weights inferred by a trained group of agents in the Collision Avoidance task.  In this task, agents are rewarded for reaching an assigned goal location while avoiding collisions.  Agents are divided into three teams, consisting of agents 1-8, 9-16, and 17-24, and are only penalized for colliding with other agents on their team.  We therefore expect agents to assign large attention weights only to other agents on their same team, because each agents' reward is independent of the actions of agents on other teams.  Fig.~\ref{fig:relevant_set_viz} displays the average attention weights as an M x M grid, with the \(i\)th row and \(j\)th column corresponding to the average attention weight that agent \(i\) assigns to agent \(j\).  Because agents always assign an attention weight of \(1\) to themselves, we remove these elements from the visualization as they are uninformative.  We find that, as expected, agents assign considerably non-zero attention weights only to other agents on their same team, while assigning near-zero attention weights to all other agents.  Attention weights were averaged over 5000 independent episodes. 

\subsection{Policy Gradient Estimator Variance Analysis}
\begin{figure}[!ht]
        \centering
        \includegraphics[width=\linewidth]{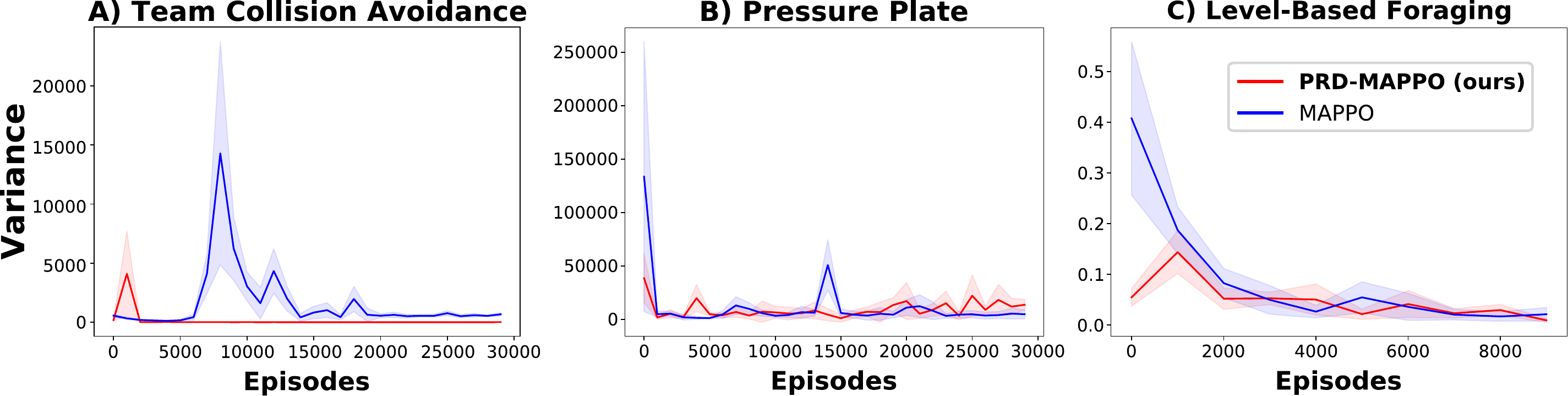}
        \vspace{-5mm}
        \caption{\textbf{Gradient estimator variance vs. episode for team collision avoidance, pressure plate, and LBF environments}.  Solid lines indicate the average over 5 random seeds, and shaded regions denote a 95\% confidence interval.  PRD-MAPPO tends to avoid the dramatic spikes in gradient variance demonstrated by MAPPO.}   
        \label{fig:gradvar}
    \end{figure}

To empirically verify the claim that partial reward decoupling decreases the variance of MAPPO policy gradient estimates, we estimate the variance of MAPPO and PRD-MAPPO at various points during training.  For maximum comparability, we compute the variance for both MAPPO and PRD-MAPPO using data gathered from the same policy, taken at 1000-episode intervals during the training of PRD-MAPPO.  Using these policies, we collect 100 independent batches of data, and differentiate the MAPPO or PRD-MAPPO surrogate objective evaluated on each batch, to obtain 100 independent gradient estimates for both approaches for each policy.  Finally, we arrive at a scalar empirical variance estimate, by taking the trace of the covariance matrix estimated using each batch of 100 gradient estimates, along with a 95\% confidence interval.  The results are plotted in Fig. \ref{fig:gradvar}.  In general, we find that PRD-MAPPO tends to avoid the spikes in gradient variance present in MAPPO, which may explain its improved stability and asymptotic performance.

\section{Related Work}

 Many recent approaches have been proposed to deal with the credit assignment problem. G2ANet \citep{g2anet}, for instance, proposed a novel attention-based game abstraction mechanism that enables the critic to better isolate important interactions among agents, and ignore unimportant ones (although explicit decoupling is not done, as in PRD). Counterfactual Multi-Agent Policy Gradient (COMA) \citep{coma} proposed a novel \textit{counterfactual baseline} that allows each agent to more precisely determine the effect that its action had on group reward by conditioning on the actions of all other agents. COMA builds on the idea of difference rewards \citep{difference_rewards}, in which each agent uses a modified reward that compares the shared reward to a counterfactual situation in which the agent took some \textit{default action}.  
Value-decomposition actor-critics (VDAC) \citep{vdac} uses value decomposition networks \citep{vdn, qmix} as critics for credit assignment in the actor-critic framework.
 Off-policy multi-agent decomposed policy gradients \citep{dop} is another multi-agent policy-gradient algorithm that uses the idea of value decomposition, but applies it to a DDPG-style off-policy policy gradient \citep{ddpg}. Finally, Learning Implicit Credit Assignment for Cooperative Multi-Agent Reinforcement Learning (LICA) \citep{lica} implicitly addressed the credit assignment problem by representing a centralized critic as a hypernetwork, and finding an end-to-end differentiable optimization setting where the policies simultaneously improve along the joint action value gradients, thus serving as a proxy for finding optimal credit assignment strategies.

\section{Limitations}

 
 The primary limitation of PRD-MAPPO is that PRD is not guaranteed to accelerate learning in every environment, because some tasks cannot be decomposed (\textit{i.e.}, each agent's relevant set contains most or all other agents).  For example, in the traffic junction experiment, it is possible that learning is only somewhat improved by PRD because interactions among agents are too dense, making decoupling less effective.

\section{Conclusions}

We addressed the shortcomings of MAPPO, a state-of-the-art multi-agent reinforcement learning algorithm. Specifically, we hypothesized that the credit assignment problem manifests itself in policy gradient estimator variance.  Based on this hypothesis, we proposed integrating PRD into MAPPO as a strategy to improve credit assignment, yielding a new multi-agent model-free RL algorithm, PRD-MAPPO.  We demonstrated that PRD-MAPPO provides significant improvements both in learning efficiency and stability, across a diverse set of tasks, compared to both MAPPO and several state-of-the-art MARL algorithms such as QMix, LICA, and COMA.  We empirically verified the hypothesis that PRD decreases the variance of the gradient estimates of MAPPO. Finally, we visualized the relevant sets inferred by PRD, and found that it correctly grouped together agents that should cooperate. The improvements in learning speed and stability, combined with decreased gradient variance and sensible relevant set estimation, indicate that PRD, used in the context of MAPPO, provides a useful credit assignment strategy for multi-agent problems.


\bibliography{main}
\bibliographystyle{rlc}
\appendix





\section{Detail Task Descriptions}
\textbf{Collision Avoidance}: 3 teams of 8 agents each exist in a square bounded 2D region. Agents receive a reward for reaching their assigned goal location, and receive a penalty for colliding with other agents belonging to the same team.  Agents therefore need only cooperate with other agents on their team to avoid collisions. Both the agents and goals are initialized in random locations. The observation space consists of the agent's position, velocity, team ID, and goal position. The agents can take 5 possible actions that allow them to move either north, south, east, west or remain stationary. The reward function is the l2 distance between the agent position and the goal position multiplied by a scalar value of 0.1. On collision, the participating agents receive a -1 reward each. The environment terminates if all agents reach their assigned goal location or 100 timesteps run out. While training decentralized policies, relative positions of all other agents and their team ID are also included in the observation space. Episodes last a maximum of 100 timesteps. This environment was modified from the cooperative navigation environment first developed by \cite{multi-agent_ac}.  The code for this environment can be found at \texttt{https://github.com/openai/multiagent-particle-envs} (MIT License).

\textbf{Pursuit}: 8 agents exist in a 16 x 16 grid with an obstacle in the center. To receive a reward, two agents must coordinate their actions to surround randomly moving ``evader'' agents on two sides. There are 30 evaders in the environment. Each pursuer observes a 7 x 7 grid centered around itself with 3 channels, indicating the positions of walls, other agents, and evaders, respectively. Once an evader is caught, it is removed from the environment. The environment terminates when every evader has been caught, or when 500 timesteps are completed.  The environment is available in the PettingZoo MARL benchmark suite \citep{terry2021pettingzoo} at \texttt{https://pettingzoo.farama.org/environments/sisl/pursuit/} (MIT License) and was first proposed by \cite{gupta2017cooperative}.

\textbf{Pressure plate}: 6 agents exist in a grid, divided into 6 separate chambers by gates. In any given chamber, a particular agent can open the gate by standing on a special grid cell known as the pressure plate. To successfully solve the task, this agent must remain on the pressure plate until the other agents have successfully moved into the next chamber.  The goal is for one particular agent to traverse all six chambers and arrive at a goal location in the final chamber.
Each agent observes a 5x5 square around its location, with a separate channel for each type of entity in the environment (e.g., walls, pressure plates, doors, agents, and goals). The agent's (x,y) coordinates are concatenated to the end of the observation vector. The action space is discrete and has five possibilities: up, down, left, right, and remain stationary. Each agent receives rewards independent of other agents. If an agent is in the room that contains their assigned plate, their reward is the negative normalized Manhattan distance between their current position and the plate. Otherwise, their reward is the number of rooms between their current room and the room that contains their assigned plate. Episodes last a maximum of 70 timesteps.  The code for this environment is available at
\texttt{https://github.com/uoe-agents/pressureplate} (MIT License).

\textbf{Level-Based Foraging (LBF)}: Agents navigate a grid world and collect food items by cooperating with other agents. Each agent and food item is assigned a level and are randomly distributed throughout the environment.  Successfully collecting a food item of a particular level requires the sum of the levels of the agents involved to be greater than or equal to the level of the food item. Agents are rewarded based on the level of the food items they help collect, divided by their contribution (their level). Reward discounting incentivizes agents to collect all food items as quickly as possible to maximize returns. The observation space consists of the agent's position in the grid, its level, relative positions of all other agents and food items, and their levels. The agents can either move in one of the four directions, collect a food item, or do nothing. Episodes last a maximum of 70 timesteps.  The code for this environment can be found at \texttt{https://github.com/semitable/lb-foraging} (MIT License).

\textbf{Lightweight StarCraft (SMAClite)}: SMAClite is a lightweight version of the StarCraft II game engine. It is computationally less expensive relative to SC II and provides a simple ``pythonic'' framework to add custom environments and make alterations to the environment logic. The observation space consists of the relative positions, unit type, health and shield strength of the agent's allies and enemies within the field of view of the agent and the health and shield strength of itself. The agents can move in any of the 4 cardinal directions, remain stationary, or attack any of the enemy agent within its field of view. Each combat scenario is run for 100 timesteps, though agents may die before this time.  We consider three different battle scenarios, 1) 5m\_vs\_6m, where 5 agent-controlled marines battle 6 enemy marines, 2) 10m\_vs\_11m, where 10 agent-controlled marines battle 11 enemy marines, and 3) 3s5z, where 3 agent-controlled stalkers and 5 agent-controlled zealots  battle 3 enemy stalkers and 5 enemy zealots. The code for SMAClite is available at \texttt{https://github.com/uoe-agents/smaclite} (MIT License).

\section{Pseudocode}
\label{sec:pseudocode}
\begin{algorithm}
\caption{PRD-MAPPO}
\begin{algorithmic}[1]
\STATE Initialize $\theta$, the parameters for policy $\pi$, $\omega$, the parameters for state-action value critic $Q$ and $\phi$, the parameters for state value critic $V$, using orthogonal initialization (Hu et al., 2020)
\STATE Set learning rate $\alpha$
\WHILE{step $\leq$ $step_{\text{max}}$}
    \STATE set data buffer $D = \{\}$
    \FOR{$i = 1$ to \textit{batch\_size}}
        \STATE $\tau = []$ -- empty list
        \STATE initialize $h^{(1)}_{0,\pi}, \ldots, h^{(M)}_{0,\pi}$ actor RNN states
        \STATE initialize $h^{(1)}_{0,V}, \ldots, h^{(M)}_{0,V}$ state value RNN states
        \STATE initialize $h^{(1)}_{0,Q}, \ldots, h^{(M)}_{0,Q}$ state-action value RNN states
        \FOR{$t = 1$ to $T$}
            \FOR{all agents $a$}
                \STATE $u^{(a)}_t, h^{(a)}_{t,\pi} = \pi(o^{(a)}_t, h^{(a)}_{t-1,\pi}; \theta)$
            \ENDFOR
            \STATE $(q^{(1)}_t, \dots q^{(M)}_t), (h^{(1)}_{t,Q} \dots h^{(M)}_{t,Q}), W_{\text{prd}, t} = Q(s^{(1)}_t \dots s^{(M)}_t, u^{(1)}_t \dots u^{(M)}_t, h^{(1)}_{t-1,Q} \dots h^{(M)}_{t-1,Q}; \omega)$
            \STATE $(v^{(1)}_t, \dots v^{(M)}_t), (h^{(1)}_{t,V} \dots h^{(M)}_{t,V})= V(s^{(1)}_t \dots s^{(M)}_t, u^{(1)}_t \dots u^{(M)}_t, h^{(1)}_{t-1,V} \dots h^{(M)}_{t-1,V}; \phi)$ -- we mask out the actions of agent \textit{a} while calculating its state value $v^{(a)}$
            \STATE Execute actions $u_t$, observe $r_t$, $s_{t+1}$, $o_{t+1}$
            \STATE $\tau \mathrel{+}= [s_t, o_t, h_{t,\pi}, h_{t,V}, u_t, r_t, s_{t+1}, o_{t+1}]$
        \ENDFOR
        \STATE Compute relevant set $R_{1},..., R_{M}$ using $W_{\text{prd}}$
        \STATE Compute return $G_{i}$ for each agent $i=1,...,M$,  to learn the $Q$ function and total relevant-set return $\bar{G}_i = \sum_{j \in R_i} G_j$ for each agent $i$ to learn $V$ function on $\tau$ and normalize with PopArt
        \STATE Compute advantage estimate $\hat{A}^{1} ,..., \hat{A}^{M}$ via GAE on state value estimates on $\tau$, using PopArt
        \STATE Split trajectory $\tau$ into chunks of length $L$
        \FOR{$l = 0, 1, \ldots, T//L$}
            \STATE $D = D \cup (\tau[l : l + T], \hat{A}[l : l + L], G[l : l + L], \bar{G}[l : l + L])$
        \ENDFOR
    \ENDFOR
    \FOR{mini-batch $k = 1, \ldots, K$}
        \STATE $b \leftarrow$ random mini-batch from $D$ with all agent data
        \FOR{each data chunk $c$ in the mini-batch $b$}
            \STATE update RNN hidden states for $\pi$, $Q$ and $V$ from first hidden state in data chunk
        \ENDFOR
    \ENDFOR
    \STATE Adam update $\theta$ on $L(\theta)$ with data $b$
    \STATE Adam update $\omega$ on $L(\omega)$ with data $b$
    \STATE Adam update $\phi$ on $L(\phi)$ with data $b$
\ENDWHILE
\end{algorithmic}
\end{algorithm}


\section{Additional Results}

\begin{figure}[!ht]
        \centering
        \includegraphics[width=0.85\textwidth]{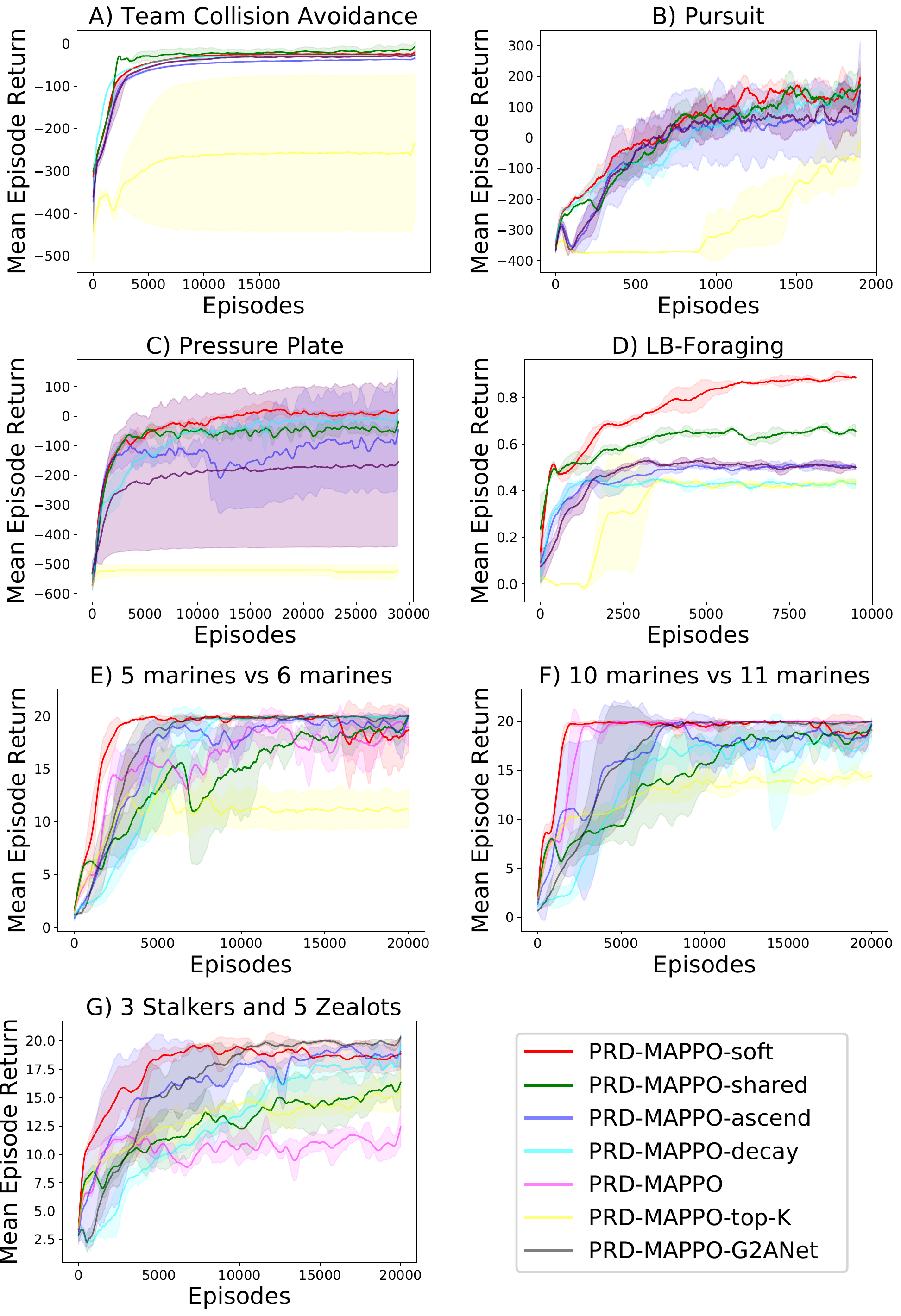}
        \caption{\textbf{Average reward vs. episode for PRD-MAPPO-soft, PRD-MAPPO-shared, PRD-MAPPO-ascend, PRD-MAPPO-decay, PRD-MAPPO, PRD-MAPPO-top-K, and PRD-MAPPO-G2ANet on A) team collision avoidance, B) pursuit, C) pressure plate, D) Level-Based Foraging tasks, E) StarCraft 5 marines vs. 6 marines, F) StarCraft 10 marines vs. 11 marines, and G) StarCraft 3 Stalkers and 5 Zealots}.  Solid lines indicate the average over 5 random seeds, and shaded regions denote a +/- 1 standard deviation confidence interval.  PRD-MAPPO-soft tended to perform the best across all tasks.}
        \label{fig:prd_variants}
\end{figure}

We experimented with various methods for selecting agent relevant sets, as described below.  Reward curves for each method in each of our four environments is shown in Fig.~\ref{fig:prd_variants}.
\textbf{PRD-MAPPO}: As described in Sec. 3.1 of the manuscript.  The attention-weight threshold \(\epsilon\) used to agent relevant sets is held constant through training.

\textbf{PRD-MAPPO-soft}: As described in Sec. 4 of the manuscript.  A variant of PRD-MAPPO in which advantage terms are not excluded from the PPO update according to hard thresholding, but rather advantage terms for each agent \(i\) are softly re-weighted according to the attention weights applied by other agents to the actions of agent \(i\).

\textbf{PRD-MAPPO-ascend}: Attention-weight threshold \(\epsilon\) is linearly increased from 0 to \(\theta\) over the first \(N\) policy updates and then held constant, where \(\theta\) and \(N\) are hyperparameters. This method transitions from including all agents in the relevant set to having only a subset of agents in the relevant set.

\textbf{PRD-MAPPO-decay}: Attention-weight threshold \(\epsilon\) is linearly decreased from \(theta\) to 0 over the first \(N\) policy updates, and then held constant.  In this case, agents aggressively prune relevant sets early on, transitioning to standard MAPPO by the end of training.

\textbf{PRD-MAPPO-G2ANet}: A semi-hard attention mechanism based on G2ANet \cite{g2anet} is used to select relevant sets.  Agents are excluded from the relevant set if their associated attention weight is exactly \(0\).  This approach has the advantage that it allows a manual threshold on attention weights to be avoided.

\textbf{PRD-MAPPO-top-k}: The agents with the top \(k\) highest attention weights are included in the relevant set (where \(k\) is a hyperparameter).

\section{Implementation Details}

The code was run on Lambda Labs deep learning workstation with 2-4 Nvidia RTX 2080 Ti graphics cards.  Each training run was run on one single GPU, and required approximately 2 days.  The hyperparamers used for our experiments are reported in the tables below:

\section{Hyperparameters}
 Hyperparameters used for MAPPO variants, PRD variants, PRD\_V\_MAPPO, QMix, LICA and COMA that are common to all tasks are shown in Tables 
 \ref{tab:common_hyperparams}\ref{tab:mappo_common_hyperparams}, 
 \ref{tab:qmix_common_hyperparams} \ref{tab:lica_common_hyperparams}, and \ref{tab:coma_common_hyperparams} respectively.  The task-specific hyperparameters considered in our grid search for MAPPO variants, PRD variants, PRD\_V\_MAPPO\, QMix, LICA, and COMA are shown in Tables \ref{table:mappo_hyperparameter_sweep}, \ref{table:prd_hyperparameter_sweep}, \ref{table:prd_old_hyperparameter_sweep} \ref{tab:qmix_hyperparam_sweep}, \ref{tab:lica_hyperparam_sweep}, and \ref{tab:coma_hyperparam_sweep}, respectively.  Bold values indicate the optimal hyperparameters.

\begin{table}[H]
\centering
\caption{Episodic Length of all environments}
\begin{tabular}{|c|c|}
\hline
\textbf{\begin{tabular}[c]{@{}c@{}}common\\ environment\end{tabular}} & \textbf{max timesteps} \\ \hline
collision avoidance & 100 \\ \hline
pursuit             & 500 \\ \hline
pressure plate      & 70 \\ \hline
level-based foraging & 70 \\ \hline
5m\_vs\_6m          & 100 \\ \hline
10m\_vs\_11m        & 100 \\ \hline
3s5z                & 100 \\ \hline

\end{tabular}

\label{tab:environment_timesteps}
\end{table}

\begin{table}[H]
\centering
\caption{Common Hyperparameters for all algorithms in all domains}
\begin{tabular}{|c|c|}
\hline
\textbf{\begin{tabular}[c]{@{}c@{}}common\\ hyperparameters\end{tabular}} & \textbf{value} \\ \hline
optimizer          & AdamW \\ \hline
gamma              & 0.99 \\ \hline
gae lambda         & 0.95 \\ \hline
weight decay       & 0.0 \\ \hline
optim epsilon      & 1e-5 \\ \hline
max grad norm      & 10.0 \\ \hline
network initialization & orthogonal \\ \hline

\end{tabular}

\label{tab:common_hyperparams}
\end{table}

\begin{table}[H]
\centering
\caption{Common Hyperparameters for MAPPO, HAPPO, MAPPO-G2ANet, PRD-V-MAPPO, PRD-MAPPO-shared and PRD-MAPPO-soft algorithms in all domains}
\begin{tabular}{|c|c|}
\hline
\textbf{\begin{tabular}[c]{@{}c@{}}common\\ hyperparameters\end{tabular}} & \textbf{value} \\ \hline
critic loss        & huber loss \\ \hline
huber delta        & 10.0 \\ \hline
num mini-batch     & 1 \\ \hline
gae lambda         & 0.95 \\ \hline 
actor network      & rnn \\ \hline
recurrent data chunk length & 10 \\ \hline
recurrent num layers & 1 \\ \hline
rnn hidden dim     & 64 \\ \hline
value normalization & PopArt \\ \hline

\end{tabular}

\label{tab:mappo_common_hyperparams}
\end{table}

\begin{table}[H]
\centering
\caption{Common Hyperparameters for QMix in all domains.}
\begin{tabular}{|c|c|}
\hline
\textbf{\begin{tabular}[c]{@{}c@{}}common\\ hyperparameters\end{tabular}} & \textbf{value} \\ \hline
buffer size        & 5000 \\ \hline
batch size         & 32    \\ \hline
hypernet layers    & 2 \\ \hline
hypernet hidden dim & 32 \\ \hline
target network update interval & 200 \\ \hline
td lambda           & 0.8 \\ \hline
epsilon decay steps & 2000 episodes \\ \hline
epsilon start       & 1.0 \\ \hline
epsilon end         & 0.1 \\ \hline
value loss          & huber loss \\ \hline
huber delta         & 10.0 \\ \hline
q network           & rnn \\ \hline
rnn hidden dim      & 64 \\ \hline
recurrent data chunk length & 10 \\ \hline
recurrent num layers & 1 \\ \hline

\end{tabular}
\label{tab:qmix_common_hyperparams}
\end{table}

\begin{table}[H]
\centering
\caption{Common Hyperparameters for LICA.}
\begin{tabular}{|c|c|}
\hline
\textbf{\begin{tabular}[c]{@{}c@{}}common\\ hyperparameters\end{tabular}} & \textbf{value} \\ \hline
hypernet layers    & 2 \\ \hline
hypernet hidden dim & 64 \\ \hline
target network update interval & 200 \\ \hline
td lambda           & 0.8 \\ \hline
critic loss         & huber loss \\ \hline
huber delta         & 10.0 \\ \hline
actor network       & rnn \\ \hline
actor rnn hidden dim & 64 \\ \hline
actor recurrent data chunk length & 10 \\ \hline
actor recurrent num layers & 1 \\ \hline

\end{tabular}
\label{tab:lica_common_hyperparams}
\end{table}

\begin{table}[H]
\centering
\caption{Common Hyperparameters for COMA.}
\begin{tabular}{|c|c|}
\hline
\textbf{\begin{tabular}[c]{@{}c@{}}common\\ hyperparameters\end{tabular}} & \textbf{value} \\ \hline
target network update interval & 200 \\ \hline
td lambda           & 0.8 \\ \hline
critic loss         & huber loss \\ \hline
huber delta         & 10.0 \\ \hline
actor network       & rnn \\ \hline
rnn hidden dim      & 64 \\ \hline
recurrent data chunk length & 10 \\ \hline
recurrent num layers & 1 \\ \hline

\end{tabular}
\label{tab:coma_common_hyperparams}
\end{table}

\begin{table}[H]
\centering
\caption{MAPPO and MAPPO-G2ANet hyperparameter sweep. Bold values indicate the optimal hyperparameters.}
{\tiny
\begin{tabular}{|c|c|c|c|c|c|c|c}
\cline{1-8}
\textbf{\begin{tabular}[c]{@{}c@{}}Environment \\ Name\end{tabular}} &
  epochs &
  num\_episodes &
  value\_lr &
  policy\_lr &
  clip &
  entropy\_pen &
   \\ \cline{1-7}
\begin{tabular}[c]{@{}c@{}}Collision\\ Avoidance\end{tabular} &
  {[}\textbf{5}, 10, 15{]} &
  {[}5, \textbf{10}{]} &
  {[}1e-4, \textbf{5e-4}, 1e-3{]} &
  {[}1e-4, \textbf{5e-4}, 1e-3{]} &
  {[}\textbf{0.05}, 0.2{]} &
  {[}1e-3, \textbf{8e-3}, 1e-2{]} &
   \\ \cline{1-7}
Pursuit &
  {[}\textbf{5}, 10, 15{]} &
  {[}2, \textbf{5}, 10{]} &
  {[}1e-4, \textbf{5e-4}, 1e-3{]} &
  {[}1e-4, \textbf{5e-4}, 1e-3{]} &
  {[}0.05, \textbf{0.2}{]} &
  {[}1e-3, 8e-3, \textbf{1e-2}{]} &
   \\ \cline{1-7}
\begin{tabular}[c]{@{}c@{}}Pressure\\ Plate\end{tabular} &
  {[}\textbf{5}, 10, 15{]} &
  {[}5, 7, \textbf{10}{]} &
  {[}1e-4, \textbf{5e-4}, 1e-3{]} &
  {[}1e-4, \textbf{5e-4}, 1e-3{]} &
  {[}0.05, 0.1, \textbf{0.2}{]} &
  {[}1e-3, \textbf{1e-2}, 5e-2, 1e-1{]} &
   \\ \cline{1-7}
\begin{tabular}[c]{@{}c@{}}Level-Based\\ Foraging\end{tabular} &
  {[}1, \textbf{5}, 10{]} &
  {[}1, 5, \textbf{10}{]} &
  {[}\textbf{5e-4}, 1e-3, 5e-3{]} &
  {[}\textbf{5e-4}, 1e-3, 5e-3{]} &
  {[}0.1, \textbf{0.2}{]} &
  {[}1e-3, 5e-3, \textbf{1e-2}{]} &
   \\ \cline{1-7}
\begin{tabular}[c]{@{}c@{}}5m\_vs\_6m\end{tabular} &
  {[}1, \textbf{5}, 10{]} &
  {[}5, \textbf{10}{]} &
  {[}1e-4, 3e-4, \textbf{5e-4}{]} &
  {[}1e-4, 3e-4, \textbf{5e-4}{]} &
  {[}0.1, \textbf{0.2}{]} &
  {[}0.0, 5e-3, \textbf{1e-2}{]} &
   \\ \cline{1-7}
\begin{tabular}[c]{@{}c@{}}10m\_vs\_11m\end{tabular} &
  {[}1, \textbf{5}, 10{]} &
  {[}5, \textbf{10}{]} &
  {[}1e-4, 3e-4, \textbf{5e-4}{]} &
  {[}1e-4, 3e-4, \textbf{5e-4}{]} &
  {[}0.1, \textbf{0.2}{]} &
  {[}0.0, 5e-3, \textbf{1e-2}{]} &
   \\ \cline{1-7}
\begin{tabular}[c]{@{}c@{}}3s5z\end{tabular} &
  {[}1, \textbf{5}, 10{]} &
  {[}5, \textbf{10}{]} &
  {[}1e-4, 3e-4, \textbf{5e-4}{]} &
  {[}1e-4, 3e-4, \textbf{5e-4}{]} &
  {[}0.1, \textbf{0.2}{]} &
  {[}0.0, 5e-3, \textbf{1e-2}{]} &
   \\ \cline{1-7}
\end{tabular}

}
\label{table:mappo_hyperparameter_sweep}
\end{table}

\begin{table}[H]
\centering
\caption{PRD-MAPPO-global and PRD-MAPPO-soft hyperparameter sweep. Bold values indicate the optimal hyperparameters.}
{\tiny
\begin{tabular}{|c|c|c|c|c|c|c|c}
\cline{1-7}
\textbf{\begin{tabular}[c]{@{}c@{}}Environment \\ Name\end{tabular}} &
  epochs &
  num\_episodes &
  value\_lr &
  policy\_lr &
  clip &
  entropy\_pen &
   \\ \cline{1-7}
\begin{tabular}[c]{@{}c@{}}Collision\\ Avoidance\end{tabular} &
  {[}\textbf{5}, 10, 15{]} &
  {[}5, \textbf{10}{]} &
  {[}1e-4, \textbf{5e-4}, 1e-3{]} &
  {[}1e-4, \textbf{5e-4}, 1e-3{]} &
  {[}\textbf{0.05}, 0.2{]} &
  {[}0.0, \textbf{1e-3}, 8e-3{]} &
   \\ \cline{1-7}
Pursuit &
  {[}\textbf{5}, 10, 15{]} &
  {[}2, \textbf{5}{]} &
  {[}1e-4, \textbf{5e-4}, 1e-3{]} &
  {[}1e-4, \textbf{5e-4}, 1e-3{]} &
  {[}0.05, \textbf{0.2}{]} &
  {[}\textbf{1e-3}, 8e-3, 1e-2{]} &
   \\ \cline{1-7}
\begin{tabular}[c]{@{}c@{}}Pressure\\ Plate\end{tabular} &
  {[}\textbf{5}, 10, 15{]} &
  {[}5, 7, \textbf{10}{]} &
  {[}1e-4, \textbf{5e-4}, 1e-3{]} &
  {[}1e-4, \textbf{5e-4}, 1e-3{]} &
  {[}0.1, \textbf{0.2}{]} &
  {[}\textbf{1e-3}, 1e-2, 5e-2, 1e-1{]} &
   \\ \cline{1-7}
\begin{tabular}[c]{@{}c@{}}Level-Based\\ Foraging\end{tabular} &
  {[}1, \textbf{5}, 10{]} &
  {[}1, 5, \textbf{10}{]} &
  {[}\textbf{5e-4}, 1e-3, 5e-3{]} &
  {[}\textbf{5e-4}, 1e-3, 5e-3{]} &
  {[}0.1, \textbf{0.2}{]} &
  {[}0.0, \textbf{1e-3}, 8e-3{]} &
   \\ \cline{1-7}
\begin{tabular}[c]{@{}c@{}}5m\_vs\_6m\end{tabular} &
  {[}1, \textbf{5}, 10{]} &
  {[}5, \textbf{10}{]} &
  {[}1e-4, 3e-4, \textbf{5e-4}{]} &
  {[}1e-4, 3e-4, \textbf{5e-4}{]} &
  {[}0.1, \textbf{0.2}{]} &
  {[}0.0, \textbf{1e-3}, 1e-2{]} &
   \\ \cline{1-7}
\begin{tabular}[c]{@{}c@{}}10m\_vs\_11m\end{tabular} &
  {[}1, \textbf{5}, 10{]} &
  {[}5, \textbf{10}{]} &
  {[}1e-4, 3e-4, \textbf{5e-4}{]} &
  {[}1e-4, 3e-4, \textbf{5e-4}{]} &
  {[}0.1, \textbf{0.2}{]} &
  {[}0.0, \textbf{1e-3}, 1e-2{]} &
   \\ \cline{1-7}
\begin{tabular}[c]{@{}c@{}}3s5z\end{tabular} &
  {[}1, \textbf{5}, 10{]} &
  {[}5, \textbf{10}{]} &
  {[}1e-4, 3e-4, \textbf{5e-4}{]} &
  {[}1e-4, 3e-4, \textbf{5e-4}{]} &
  {[}0.1, \textbf{0.2}{]} &
  {[}0.0, \textbf{1e-3}, 1e-2{]} &
   \\ \cline{1-7}
\end{tabular}

}
\label{table:prd_hyperparameter_sweep}
\end{table}

\begin{table}[H]
\centering
\caption{PRD-V-MAPPO hyperparameter sweep. Bold values indicate the optimal hyperparameters.}
{\tiny
\begin{tabular}{|c|c|c|c|c|c|c|c|c}
\cline{1-8}
\textbf{\begin{tabular}[c]{@{}c@{}}Environment \\ Name\end{tabular}} &
  epochs &
  num\_episodes &
  value\_lr &
  policy\_lr &
  clip &
  entropy\_pen &
  threshold &
   \\ \cline{1-8}
\begin{tabular}[c]{@{}c@{}}Collision\\ Avoidance\end{tabular} &
  {[}\textbf{5}, 10, 15{]} &
  {[}5, \textbf{10}{]} &
  {[}1e-4, \textbf{5e-4}, 1e-3{]} &
  {[}1e-4, \textbf{5e-4}, 1e-3{]} &
  {[}\textbf{0.05}, 0.2{]} &
  {[}0.0, 1e-3, \textbf{1e-2}{]} &
  {[}0.05, \textbf{0.12}, 0.2{]} &
   \\ \cline{1-8}
Pursuit &
  {[}\textbf{5}, 10, 15{]} &
  {[}2, \textbf{5}{]} &
  {[}1e-4, \textbf{5e-4}, 1e-3{]} &
  {[}\textbf{1e-4}, 5e-4, 1e-3{]} &
  {[}0.05, \textbf{0.2}{]} &
  {[}1e-3, 8e-3, \textbf{1e-2}{]} &
  {[}0.2, \textbf{0.3}, 0.5{]} &
   \\ \cline{1-8}
\begin{tabular}[c]{@{}c@{}}Pressure\\ Plate\end{tabular} &
  {[}\textbf{5}, 10, 15{]} &
  {[}5, 7,  \textbf{10}{]} &
  {[}1e-4, \textbf{5e-4}, 1e-3{]} &
  {[}\textbf{1e-4}, 5e-4, 1e-3{]} &
  {[}0.1, \textbf{0.2}{]} &
  {[}1e-3, 1e-2, \textbf{5e-2}, 1e-1{]} &
  {[}\textbf{0.2}, 0.4{]} &
   \\ \cline{1-8}
\begin{tabular}[c]{@{}c@{}}Level-Based\\ Foraging\end{tabular} &
  {[}1, \textbf{5}, 10{]} &
  {[}1, 5, \textbf{10}{]} &
  {[}\textbf{5e-4}, 1e-3, 5e-3{]} &
  {[}\textbf{5e-4}, 1e-3, 5e-3{]} &
  {[}0.1, \textbf{0.2}{]} &
  {[}0.0, 1e-3, \textbf{8e-3}{]} &
  {[}0.15, \textbf{0.2}, 0.33{]} &
   \\ \cline{1-8}
\begin{tabular}[c]{@{}c@{}}5m\_vs\_6m\end{tabular} &
  {[}1, \textbf{5}, 10{]} &
  {[}5, \textbf{10}{]} &
  {[}1e-4, 3e-4, \textbf{5e-4}{]} &
  {[}1e-4, 3e-4, \textbf{5e-4}{]} &
  {[}0.1, \textbf{0.2}{]} &
  {[}0.0, 5e-3, \textbf{1e-2}{]} &
  {[}0.15, \textbf{0.2}, 0.33{]} &
   \\ \cline{1-8}
\begin{tabular}[c]{@{}c@{}}10m\_vs\_11m\end{tabular} &
  {[}1, \textbf{5}, 10{]} &
  {[}5, \textbf{10}{]} &
  {[}1e-4, 3e-4, \textbf{5e-4}{]} &
  {[}1e-4, 3e-4, \textbf{5e-4}{]} &
  {[}0.1, \textbf{0.2}{]} &
  {[}0.0, 5e-3, \textbf{1e-2}{]} &
  {[}\textbf{0.1}, 0.2, 0.33{]} &
   \\ \cline{1-8}
\begin{tabular}[c]{@{}c@{}}3s5z\end{tabular} &
  {[}1, \textbf{5}, 10{]} &
  {[}5, \textbf{10}{]} &
  {[}1e-4, 3e-4, \textbf{5e-4}{]} &
  {[}1e-4, 3e-4, \textbf{5e-4}{]} &
  {[}0.1, \textbf{0.2}{]} &
  {[}0.0, 5e-3, \textbf{1e-2}{]} &
  {[}\textbf{0.12}, 0.2, 0.33{]} &
   \\ \cline{1-8}
\end{tabular}

}
\label{table:prd_old_hyperparameter_sweep}
\end{table}

\begin{table}[H]
\centering
\scriptsize
\caption{Hyperparameter sweep for QMix. Bold values were selected for training the agent.}
\label{tab:qmix_hyperparam_sweep}
\begin{tabular}{|c|c|c|c|}
\hline
\textbf{\begin{tabular}[c]{@{}c@{}}Environment\\ Name\end{tabular}} &
  \multicolumn{1}{l|}{learning rate} &
  \multicolumn{1}{l|}{update interval (episodes)} &
  \multicolumn{1}{l|}{hard interval} \\ \hline
\begin{tabular}[c]{@{}c@{}}Collision\\ Avoidance\end{tabular} & {[}1e-4, \textbf{5e-4}, 1e-3{]} & {[}5, \textbf{10}, 20{]} & {[}100, \textbf{200}, 500{]} \\ \hline
Pursuit                                                       & {[}1e-4, \textbf{5e-4}, 1e-3{]} & {[}\textbf{5}, 10, 20{]} & {[}100, \textbf{200}, 500{]} \\ \hline
\begin{tabular}[c]{@{}c@{}}Pressure\\ Plate\end{tabular}      & {[}1e-4, \textbf{5e-4}, 1e-3{]} & {[}5, \textbf{10}, 20{]} & {[}100, \textbf{200}, 500{]} \\ \hline
LB-Foraging                                                   & {[}1e-4, \textbf{5e-4}, 1e-3{]} & {[}5, \textbf{10}, 20{]} & {[}100, \textbf{200}, 500{]} \\ \hline
5m\_vs\_6m                                                   & {[}1e-4, \textbf{5e-4}, 1e-3{]} & {[}5, \textbf{10}, 20{]} & {[}100, \textbf{200}, 500{]} \\ \hline
10m\_vs\_11m                                                   & {[}1e-4, \textbf{5e-4}, 1e-3{]} & {[}5, \textbf{10}, 20{]} & {[}100, \textbf{200}, 500{]} \\ \hline
3s5z                                                   & {[}1e-4, \textbf{5e-4}, 1e-3{]} & {[}5, \textbf{10}, 20{]} & {[}100, \textbf{200}, 500{]} \\ \hline
\end{tabular}
\end{table}

\begin{table}[H]
\centering
\caption{Hyperparameter sweep for LICA. Bold values were selected for training the agent.}
\begin{tabular}{|c|c|c|c|}
\hline
\textbf{\begin{tabular}[c]{@{}c@{}}Environment\\ Name\end{tabular}} &
  \multicolumn{1}{l|}{critic\_lr} &
  \multicolumn{1}{l|}{actor\_lr} &
  \multicolumn{1}{l|}{entropy\_coeff} \\ \hline
\begin{tabular}[c]{@{}c@{}}Collision\\ Avoidance\end{tabular} & {[}1e-4, \textbf{5e-4}, 1e-3{]} & {[}\textbf{1e-4}, 5e-4, 1e-3{]} & {[}1e-2, \textbf{1e-1}{]} \\ \hline
    Pursuit                                                        & {[}\textbf{1e-4}, 5e-4, 1e-3{]} & {[}\textbf{1e-4}, 5e-4, 1e-3{]} & {[}1e-2, \textbf{1e-1}{]} \\ \hline
\begin{tabular}[c]{@{}c@{}}Pressure\\ Plate\end{tabular}       & {[}\textbf{1e-4}, 5e-4, 1e-3{]} & {[}\textbf{1e-4}, 5e-4, 1e-3{]} & {[}1e-2, \textbf{1e-1}{]} \\ \hline
LB-Foraging                                                    & {[}1e-3, \textbf{5e-3}, 1e-2{]} & {[}\textbf{1e-3}, 5e-3, 1e-2{]} & {[}1e-2, \textbf{1e-1}{]} \\ \hline
5m\_vs\_6m                                                   & {[}1e-4, \textbf{5e-4}, 1e-2{]} & {[}1e-4, \textbf{5e-4}, 1e-3{]} & {[}1e-2, \textbf{1e-1}{]} \\ \hline
10m\_vs\_11m                                                   & {[}1e-4, \textbf{5e-4}, 1e-2{]} & {[}1e-4, \textbf{5e-4}, 1e-3{]} & {[}1e-2, \textbf{1e-1}{]} \\ \hline
3s5z                                                   & {[}1e-4, \textbf{5e-4}, 1e-2{]} & {[}1e-4, \textbf{5e-4}, 1e-3{]} & {[}1e-2, \textbf{1e-1}{]} \\ \hline
\end{tabular}
\label{tab:lica_hyperparam_sweep}
\end{table}

\begin{table}[H]
\centering
\caption{Hyperparameter sweep for COMA. Bold values indicate the optimal hyperparameters.}
\begin{tabular}{|c|c|c|c|}
\hline
\textbf{\begin{tabular}[c]{@{}c@{}}Environment\\ Name\end{tabular}} &
  \multicolumn{1}{l|}{value\_lr} &
  \multicolumn{1}{l|}{policy\_lr} &
  \multicolumn{1}{l|}{entropy\_coeff} \\ \hline
\begin{tabular}[c]{@{}c@{}}Collision\\ Avoidance\end{tabular} &
  {[}1e-4, 5e-4, \textbf{1e-3}{]} &
  {[}5e-4, \textbf{7e-4}, 1e-3{]} &
  {[}1e-3, 8e-3, \textbf{1e-2}{]} \\ \hline
Pursuit &
  {[}1e-4, \textbf{5e-4}, 1e-3{]} &
  {[}\textbf{1e-4}, 5e-4, 1e-3{]} &
  {[}1e-3, \textbf{8e-3}, 1e-2{]} \\ \hline
\begin{tabular}[c]{@{}c@{}}Pressure\\ Plate\end{tabular} &
  {[}1e-4, \textbf{5e-4}, 1e-3{]} &
  {[}\textbf{1e-4}, 5e-4, 1e-3{]} &
  {[}1e-3, \textbf{8e-3}, 1e-2{]} \\ \hline
LB-Foraging &
  {[}1e-3, \textbf{5e-3}, 1e-2{]} &
  {[}\textbf{1e-3}, 5e-3, 1e-2{]} &
  {[}1e-3, 8e-3, \textbf{1e-2}{]} \\ \hline
5m\_vs\_6m &
  {[}1e-4, \textbf{5e-4}, 1e-3{]} &
  {[}\textbf{1e-4}, 5e-4, 1e-3{]} &
  {[}1e-3, 8e-3, \textbf{1e-2}{]} \\ \hline
10m\_vs\_11m &
  {[}1e-4, \textbf{5e-4}, 1e-3{]} &
  {[}\textbf{1e-4}, 5e-4, 1e-3{]} &
  {[}1e-3, 8e-3, \textbf{1e-2}{]} \\ \hline
3s5z &
  {[}1e-4, \textbf{5e-4}, 1e-3{]} &
  {[}\textbf{1e-4}, 5e-4, 1e-3{]} &
  {[}1e-3, 8e-3, \textbf{1e-2}{]} \\ \hline
\end{tabular}
\label{tab:coma_hyperparam_sweep}
\end{table}

\end{document}